\documentclass[12pt,article,nofootinbib]{revtex4}
\usepackage{pdfpages}
\usepackage{epsf}
\usepackage{subfigure}
\usepackage{amsmath}
\usepackage{eqnarray}
\usepackage{epstopdf}
\usepackage{mathrsfs}
\usepackage{natbib}
\usepackage{amssymb}
\usepackage{dcolumn}
\usepackage{graphicx}
\usepackage{color}
\usepackage{hyperref}
\usepackage{enumerate}
\usepackage{float}

\usepackage{multirow}
\usepackage{blindtext}

\makeatletter
\newcommand*{\rom}[1]{\expandafter\@slowromancap\romannumeral #1@}
\makeatother
\def\be{\begin{equation}}
\def\ee{\end{equation}}
\def\ba{\begin{eqnarray}}
\def\ea{\end{eqnarray}}

\graphicspath{{images/}}

\begin{document}
\title{History of a Particle Bounded to the Cosmological LTB Black Hole Surrounded by the Quintom Field}
\author{Sareh Eslamzadeh}
\email{S.Eslamzadeh@stu.umz.ac.ir}
\affiliation{Department of Physics, College of Sciences, Yasouj University, 75918-74934, Yasouj, Iran.}	
\author{Saheb Soroushfar}
\thanks{Corresponding Author}
\email{Soroush@yu.ac.ir}
\affiliation{Department of Physics, College of Sciences, Yasouj University, 75918-74934, Yasouj, Iran.}

\begin{abstract}
In this paper, we derived the complete set of time-dependent geodesic equations for an LTB black hole surrounded by a Quintom field and investigated the evolution of effective potential and photon orbits across cosmic epochs. Our findings demonstrate that in an accelerated universe, the peak of an effective potential decreases in height and shifts toward smaller radii. Additionally, the probability of stable orbit formation decreases as cosmic expansion progresses. We classified the possible trajectories into four types: terminating bound orbits, stable orbits, scattering flyby orbits, and terminating escape orbits. The results indicate that stable bound orbits are more prevalent in the early universe, whereas at late time epochs, flyby orbits become dominant due to the expansion-driven weakening of gravitational potential.

We further analyzed the impact of angular momentum on the evolution of orbits, showing that as it increases, the ISCO radius decreases while the peak of the effective potential shifts outward. This suggests that particles with higher angular momentum follow extended bound orbits, and more energetic photons are more likely to be captured by the black hole. Conversely, an increase in angular momentum reduces the probability of flyby orbits while increasing the likelihood of direct fall into the black hole.

Our study provides new insights into how cosmic acceleration influences black hole geodesics, revealing that the progressive shrinking of ISCO and stable orbits eventually disappear as the universe approaches the Big Rip singularity. These findings contribute to a deeper understanding of the dynamical nature of cosmological black holes and may offer new perspectives for observational tests through gravitational lensing, accretion disk evolution, and quasi-periodic oscillations (QPOs) in evolving black hole spacetimes.
		
\end{abstract}

\maketitle
\newpage
\tableofcontents
			
	\section{Introduction}
After discovering both the stationary black hole and the accelerated universe, the issue was raised about considering a black hole in this accelerated universe. Therefore, the subject of cosmological black holes was introduced with this content. The first suggestion for a black hole in the accelerating Friedmann-Lema\^\i{}tre-Robertson-Walker (FLRW) universe can be found in McVittie’s research  \cite{33McV}. Later, other famous solutions have been suggested by Einstein and Strauss \cite{45Ein}, Vaidya \cite{77Vai}, and Lema\^\i{}tre-Tolman-Bondi (LTB) \cite{34Tol,47Bon,97Lem}. The obvious characteristic of all these solutions is the dynamical component existence such as time in the black hole metric. Therefore, in all solutions, the basic definitions of stationary black holes such as mass, surface gravity, event horizon, etc., have been redefined based on the existence of black holes in a dynamical background \cite{94Hay, 02Ash, 14Far}. In addition, we notice that each solution answers part of the relevant questions and leaves some ambiguity; in Ref. \cite{15Far}, one can find a helpful review on the cosmological black hole. In this research, we pay specific attention to the LTB solution which has been probed in various research like Refs. \cite{10Fir, 11Gao, 11Chak, 22Kok, 23Sch}. In Ref. \cite{10Fir}, authors have introduced the LTB black hole as a collapsing over-dense region extending to an expanding closed, open, or flat asymptotically FLRW universe. In Ref. \cite{11Gao}, the metric of LTB black hole has been generalized to the LTB black hole in the several backgrounds of a spatially flat, dark matter, dark matter plus dark energy, or dark energy dominated universe. In Ref. \cite{11Chak}, three laws of thermodynamics for the inhomogeneous spherically symmetric LTB model has been analyzed. Besides, in Ref. \cite{22Kok}, light-rays propagating through the LTB geometrical structure has been investigated. Also, in the newest research in Ref. \cite{23Sch}, the evolution of spherical symmetric inhomogeneities of $\Lambda$LTB model have been compared with inhomogeneous cosmological models based on $f(R)$ modified gravity theories.\\
		
The most remarkable aspect of the accelerated universe is how the acceleration has been created. On the one hand, the answer of particle physics to this question is the existence of some fields like Quintessence \cite{13Tsu},  Phantom \cite{02Cal}, K-essence \cite{01Arm}, and Tachyon \cite{02Sen} as a Dark Energy. On the other hand, observational data illustrate the transition for the value of the equation of state from $w>-1$ at the prior epochs to $w<-1$ at the late epochs of the cosmic time. Therefore, it is undeniable that we pay special attention to the field that expects this transition. It can be considered the combination of two field components like Quintessence and Phantom such that the evolution of kinetic energy of those two fields describes $w$ crosses $-1$. Such a field is named Quintom which has been considered in recent research \cite{04Eliz,05Fen,08Noz,10Set,16Noz}. Also, Ref. \cite{08Eli} provides a helpful demonstration of the universe's history by comparing single or multiple scalar field theories. Therefore, we can interpret that there are cosmological LTB black holes surrounded by the field that accelerates the universe. Likewise, Refs. \cite{11Gao, 08Gao, 17Lei, 23Far} can be mentioned as instance research in this issue. Moreover, we are also interested in other aspects of the massive objects in the accelerated universe such as the geodesics of spacetime which can provide a perfect understanding of the motion of light rays in all time cosmic epochs. Indeed, attention to the geodesics equation of a gravitational field is an essential feature for predicting observational effects like gravitational time delay, light deflection, and perihelion shift. In this regard, we have provided a comprehensive investigation of the geodesics and possible orbits in Refs. \cite{15Sor,16Sor, 22Sor}. Especially, in Ref. \cite{16Sor}, we have probed in the geodesics of the Kerr-Newman-AdS spacetime numerically. As well, in Ref. \cite{22Sor}, we have done the analytical research on the black hole immersed in dust, radiation, cosmological constant, and Phantom field, separately. Besides, it has been investigated in Refs. \cite{14Nol, 16Ant} for the McVittie black hole as a cosmological black hole. Also, in Ref. \cite{02Bak}, applying the Lema\^\i{}tre-Schwarzschild solution, the author has achieved domains of attraction for some local structures like the Milky Way or Virgo Cluster. In addition, authors in Ref. \cite{04Nes} have done interesting numerical research about bound systems like the Milky Way in the universes with accelerating expansion either increases with time for phantom cosmologies or decreases with time for quintessence.
		
In this paper, we review the cosmological LTB black hole surrounded by the Quintom field in Section II. Afterward, we investigate the effective potential and photon orbit for a particle bounded to such a black hole in the accelerated universe with Quintom content in section III. We provide conclusion in section IV.

	\section{Outlook on the Cosmological LTB Black Hole with the Quintom field as a Content of the Universe }
	Cosmological LTB black hole was introduced by Lema\^\i{}tre, then it was investigated by Tolman and Bondi. So, such a black hole in the FRW universe has formed entitled LTB black hole \cite{00Cel} with the line element as the form below 
	  \begin{equation}\label{LTB}
	ds^2=dt^2-\frac{R'^2 (r,t)}{1+2 E(r)} dr^2-R^2 (r,t) (d\theta^2+\sin^2 \theta d\varphi^2).
	\end{equation}
In this metric, $R(r,t)$ is a physical radius and, 
	\begin{equation}\label{E-tot}
E(r)=\frac{1}{2} \dot{R}^2(r,t) -\frac{M(R)}{R(r,t)},
	\end{equation}
is the interpretation of the total energy per unit mass, where $M(R)$ could be Misner-Sharp mass as follows
	\begin{equation}\label{Mass}
M_{M-Sh}=M+\frac{4\pi}{3} \rho R^3.
	\end{equation}
Also, we notice that hereafter overdot and prime denote differentiation with respect to $t$ and $r$, respectively. Afterwards, one can considers a perfect fluid like Quintom as a content of the universe with the energy-momentum tensor as follows
	\begin{equation}\label{E-MT}
	T_{\mu\nu}=(\rho_q+p_q) u_{\mu} u_{\nu} + p g_{\mu\nu} ,
	\end{equation}
where $\rho_q$ and $p_q$ are density and pressure of the Quintom field, respectively; and $u^{\mu}=(1,0,0,0)$ is the four-velocity. The Quintom field has been made of these two field compositions \cite{05Fen, 10Set}. So that, the equation of state of the Quintom field has been written as follows
\begin{equation}\label{EoS1}
	w_{q}=\frac{\frac{1}{2}\dot{\delta}^2-V_{\delta}(\delta)-\frac{1}{2}\dot{\sigma}^2-V_{\sigma}(\sigma)}{\frac{1}{2}\dot{\delta}^2+V_{\delta}(\delta)-\frac{1}{2}\dot{\sigma}^2+V_{\sigma}(\sigma)},
\end{equation}
where $V_{\delta}(\delta)$ and $V_{\sigma}(\sigma)$ are the potentials for the canonical Quintessence and the Phantom field, respectively. Therefore, it is easy to conclude that the equation of state for the Quintom field crosses $-1$  in the near past because the difference in kinetic energy between the two fields evolves.
Afterward, to reconstruct the LTB black hole surrounded by the Quintom field, we take on the general line element as the form
	\begin{equation}\label{LElement}
	ds^2=dt^2-e^{\bar{\phi}(t,r)} dr^2-e^{\phi(t,r)} (d\theta^{2}+\sin^{2}\theta ~d\varphi^{2}),
	\end{equation}
where $t$ is a cosmic time parameter and $(r,\theta,\varphi)$ are co-moving coordinates. The co-moving observer finds a spatially homogeneous pressure since the source is a single perfect fluid and the background is spatially flat \cite{11Gao}. Therefore, the pressure has been assumed spatially homogeneous in the form
	\begin{equation}\label{peq}
	p_q=- \frac{p_0}{(t_{BR}-t)^2} ,
	\end{equation}
while $p_0$ is a positive constant, and $t_{BR}$ is recognized as the Big Rip singularity time. Based on the investigation and calculation  of Ref. \cite{11Gao}, the Einstein's field equations would be
	\begin{equation}\label{Sol3}
	R(r,t)\equiv x\equiv e^{\phi/2}=\bigg[r^{\frac{3}{2}}~\tau^{\frac{1-k}{2}}-\bigg(\frac{3}{2}\sqrt{2M}+\sqrt{6\pi\rho_0}  r^{\frac{3}{2}}\bigg)~\tau^{\frac{1+k}{2}}\bigg]^{\frac{2}{3}} .
	\end{equation}
while $\tau=t_{BR}-t$, and $k\equiv\sqrt{1+24\pi p_0}$ is a constant. Also, 
	\begin{equation}\label{Sol2}
	e^{\bar{\phi}}=\frac{\phi^{\prime2}}{4} e^{\phi}.
	\end{equation}
has been gained from the Einstein equations.

If we ignore the Quintom role in this metric and put $p_0=0$, the metric solution describes a black hole in the dust-dominated universe with $\rho_0=\rho_d a^3$; while, $\rho_d$ and $a$ are density and scale factor of the universe, respectively. Also, in this condition, on sufficiently large spatial scales from black hole, the LTB metric appears in homogeneous and isotropic FLRW metric. Indeed, a comoving observer can recover the FLRW limit with the scale factor $a=(1-\sqrt{6\pi\rho_0} t)^{2/3}$ as we expect from the dust-dominated universe. Hence, considering $p_0=0$ and $\rho_d=0$, the Schwarzschild solution has been recovered. In Ref. \cite{23Es}, we investigate some dynamical features, thermodynamics, and tunneling processes from the trapping horizon of this black hole. Final noticeable point is the time evolution of the cosmological and black hole apparent horizons of the LTB black hole in the Quintom universe [\cite{11Gao}, \cite{23Es}]. Indeed, in a universe dominated by Quintom dark energy, the black hole apparent horizon initially shrinks during the matter-dominated era but later expands in the phantom-dominated epoch. Conversely, the cosmological apparent horizon expands at first and then contracts. At a particular point in the past, the two horizons overlapped. As time progresses and the universe approaches the Big Rip, the horizons converge once more before ultimately vanishing. So that, two coincident apparent horizons of the black hole are created in the past, they evolve away from each other and before the Big Rip singularity, due to the existence of the Quintom component, they shrink and coincide again. We benefit from this point to the exhibition of the cosmological apparent horizon in the geodesics plots.
In the next section, we aim to calculate the geodesics and the effective potential of a mass-less particle bounded to the LTB black hole immersed in the Quintom universe.

	\section{The History of a Particle Bound to the LTB Black Hole in the Quintom Universe}
\subsection{Effective Potential}

	First, we consider the mass-less particle bounded to a cosmological black hole surrounded by the Quintom field and calculate the effective potential. Afterward, we probe the particles' total energy and effective potential in the different cosmic epochs. Thus, in the first step, we consider the general Geodesic equation as the form \cite{93Pee}
	\begin{equation}\label{Geo}
\Big(\frac{ds}{dt}\Big) \frac{d}{dt}\Big[g_{\mu \alpha}\Big(\frac{ds}{dt}\Big)^{-1} \dot{x}^{\alpha}\Big]=\frac{1}{2} g_{\alpha \beta, \mu} \dot{x}^\alpha \dot{x}^\beta,
	\end{equation}
where, $s$ is determined by Eq. (\ref{LElement}) and dot is shown as derivative with respect to time. Then, based on the generic line element has been introduced in Eq. (\ref{LElement}), we are able to obtain four geodesic equations with $\mu=0,1,2,3$, respectively, as follows
	\begin{equation}\label{mathcal A}
	\mathcal{A}=\Big(\frac{ds}{dt}\Big) \frac{d}{dt}\Big[\Big(\frac{ds}{dt}\Big)^{-1} \Big]=- \frac{1}{2}\:\Big(\dot{\bar{\phi}} \: e^{\bar{\phi}} \: \dot{r}^2+\dot{\phi}\:e^{\phi} \: (\dot{\theta}^2+\sin^2\theta \: \dot{\varphi}^2)\Big).
	\end{equation}

 	\begin{equation}\label{Geo1}
\mathcal{A} \:e^{\bar{\phi}} \: \dot{r}+ \dot{\bar{\phi}} \: e^{\bar{\phi}} \: \dot{r}+ e^{\bar{\phi}} \: \ddot{r}=\frac{1}{2}\Big(\bar{\phi}' e^{\bar{\phi}} \: \dot{r}^2+\phi' \: e^\phi \: (\dot{\theta}^2+\sin^2\theta \: \dot{\varphi}^2)\Big),
	\end{equation}
	
	\begin{equation}\label{Geo2}
\mathcal{A}\:e^{\phi} \: \dot{\theta}+ \frac{d}{dt}\Big[e^{\phi}\  \dot{\theta}\Big]= e^{\phi} \sin\theta \cos\theta \ \dot{\varphi}^2,
	\end{equation}
	
	\begin{equation}\label{Geo3}
\Big(\frac{ds}{dt}\Big) \frac{d}{dt}\Big[e^\phi \sin^2\theta \: \Big(\frac{ds}{dt}\Big)^{-1} \: \dot{\varphi}\Big]=0.
	\end{equation}
	
Paying attention to particles moving in the equatorial plane which given by considering $\theta=\frac{\pi}{2}$, from Eq. (\ref{Geo3}) we have
	\begin{equation}\label{Constant}
L=e^{\phi} \dot{\varphi}\:\Big(\frac{ds}{dt}\Big)^{-1},
	\end{equation}
where $L$ interpreting angular momentum of the particle while we deal with time dependent metric. Therefore, referring Eq. (\ref{LElement}), we have
	\begin{equation}\label{Constant2}
\dot{\varphi}^2= \frac{L^2 e^{-2\phi} (1-e^{\bar{\phi}}\: \dot{r}^2)}{1+L^2 \: e^{-\phi}}.
	\end{equation}
Finally, substituting Eqs. (\ref{mathcal A}) and (\ref{Constant2}) in Eq. (\ref{Geo1}), we calculate the $\ddot{r}$ as follows
	\begin{equation}\label{rddot}
\ddot{r}=\frac{1}{2}\dot{\bar{\phi}}\: e^{\bar{\phi}}\: \dot{r}^3+(\frac{1}{2} \bar{\phi}' \dot{r}-\dot{\bar{\phi}})\:\dot{r}+\Big(\frac{\dot{r}\dot{\phi}e^{\phi}}{2}+\frac{2}{\phi'}\Big)\Big(\frac{L^2 e^{-2\phi} (1-e^{\bar{\phi}}\: \dot{r}^2)}{1+L^2 \: e^{-\phi}}\Big).
	\end{equation}
	
Now, based on Eq. (\ref{Sol3}) we have $r=r(x,t)$, therefore we can easily gain $2\dot{x}=(\dot{\phi}+\phi' \dot{r})x$ and $2\ddot{x}=(\phi'\ddot{r} x+\dot{\phi}'\dot{r}x+\ddot{\phi}x+\phi'\dot{r}\dot{x}+\dot{\phi}\dot{x})$. Then, we rearrange the Eq. (\ref{rddot}) regarding Eqs. (\ref{Sol3}) and (\ref{Sol2}) as follows
	\begin{align}\label{xddot}
	\ddot{x}&=\Big[\frac{\ddot{\phi}}{2}+\frac{\dot{\phi}^2}{4}\Big]\:x+\bigg[\frac{2\dot{\phi}'+\dot{\phi}\phi'}{16\phi'}\bigg](2\dot{x}-x\dot{\phi})^3-\frac{L^2}{x^3}\nonumber\\
	&-\frac{L^2}{x(1+\frac{L^2}{x^2})}\bigg[\frac{\dot{\phi}}{4x}(2\dot{x}-x\dot{\phi})-\Big(\frac{2\dot{x}-x\dot{\phi}}{x}\Big)^2-\frac{\dot{\phi}}{4x}(2\dot{x}-x\dot{\phi})^3-\frac{L^2}{x^4}\bigg].
	\end{align}
It is worth noting that to derive Eq. (\ref{xddot}), we have neglected the second-order derivatives with respect to $r$. Eventually, we calculate the effective potential as follows
	\begin{align}\label{veff}
	V_{eff}&=\frac{1}{4} \bigg[x^2 \ddot{\phi} + \frac{x^2 \dot{\phi}^2}{2} + 6 L^2 x \dot{x} \dot{\phi}^3 -\frac{1}{2} L^2 x^2 \dot{\phi}^4 \nonumber\\
	&-\frac{(2 \dot{x} - x \dot{\phi})^4 (-2 \dot{\phi}' - \dot{\phi} \dot{\phi}')}{16 \dot{\phi} \phi'} -\frac{2}{L} \Big[9 L^2 \dot{x} \dot{\phi}- 4 L^2 \dot{x}^3 \dot{\phi} + 3 L^4 \dot{x} \dot{\phi}^3\Big] \arctan\Big[\frac{x}{L}\Big] \nonumber\\
	& -4 (1 - 4 \dot{x}^2) \log[x] + \frac{1}{2} (4 - 16 \dot{x}^2 + 5 L^2 \dot{\phi}^2 - 
	12 L^2 \dot{x}^2 \dot{\phi}^2 + L^4 \dot{\phi}^4) \log[L^2 + x^2]\bigg].
	\end{align}
	
\begin{figure}[ht]
		\centering
		\includegraphics[height=5 cm,width=7 cm]  {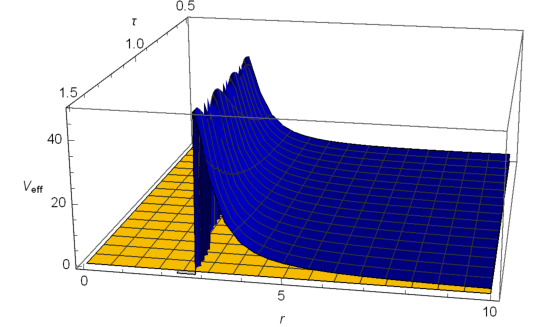}
		\hspace*{1cm}
		\includegraphics[height=5 cm,width=7 cm] {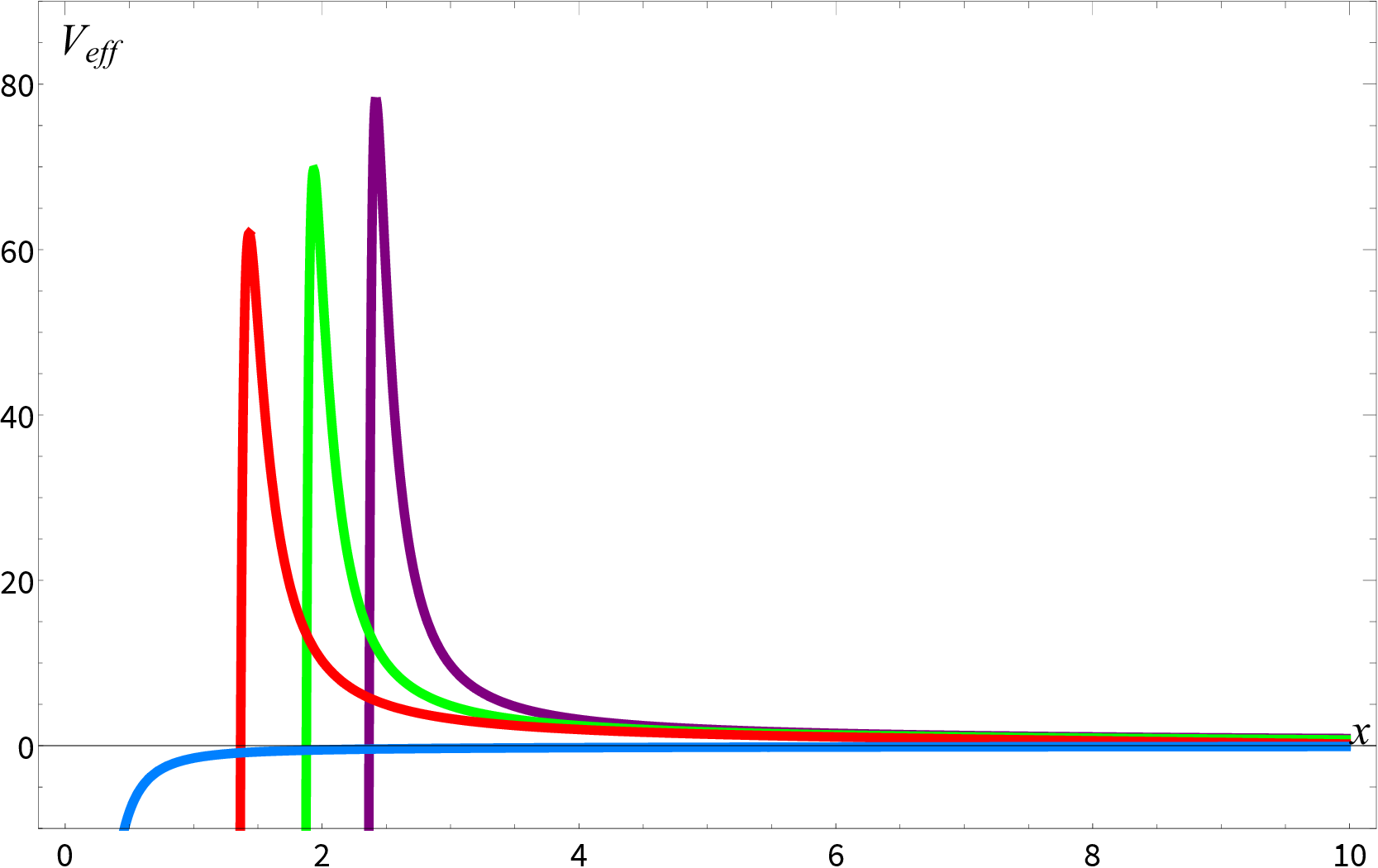}
		\caption{\scriptsize{The left-hand plot shows the behavior of the effective potential versus distance and cosmic time while the Right-hand plot shows the effective potential versus distance in some cosmic time. Plots have been depicted with different proper times: $\tau=1.4$ for the dashed-purple plot, $\tau=1$ for the green plot, and $\tau=0.6$ for red-dotted-dashed plot; Blue-Dashed plot has been shown the Schwarzschild case. Furthermore, other numerical considerations are the mass of the black hole, $M=1$, the initial angular momentum of the massless particle, $L=1.5$, $\rho_0=0.0002$, $p_0=0.001$, and $t_{BR}=1$.}}
		\label{figure_1}
	\end{figure}
This equation is a key result of our research, which we further analyze and discuss in the subsequent sections.
	
\subsection{Shape of the Orbits}
\begin{figure}[ht]
	\centering
	\includegraphics[height=4 cm,width=5 cm]  {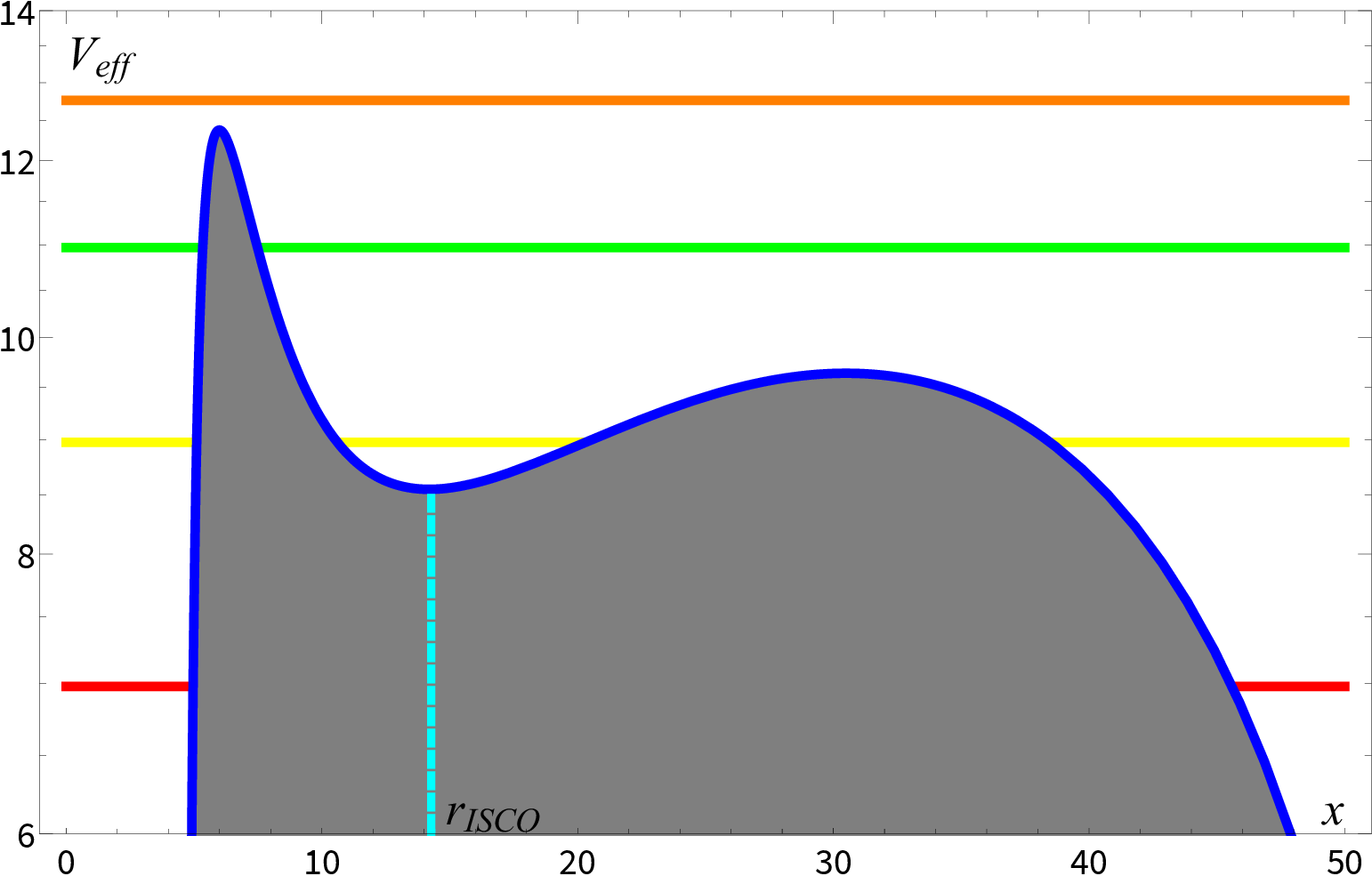}
	\hspace*{0.5cm}
	\includegraphics[height=4 cm,width=5 cm] {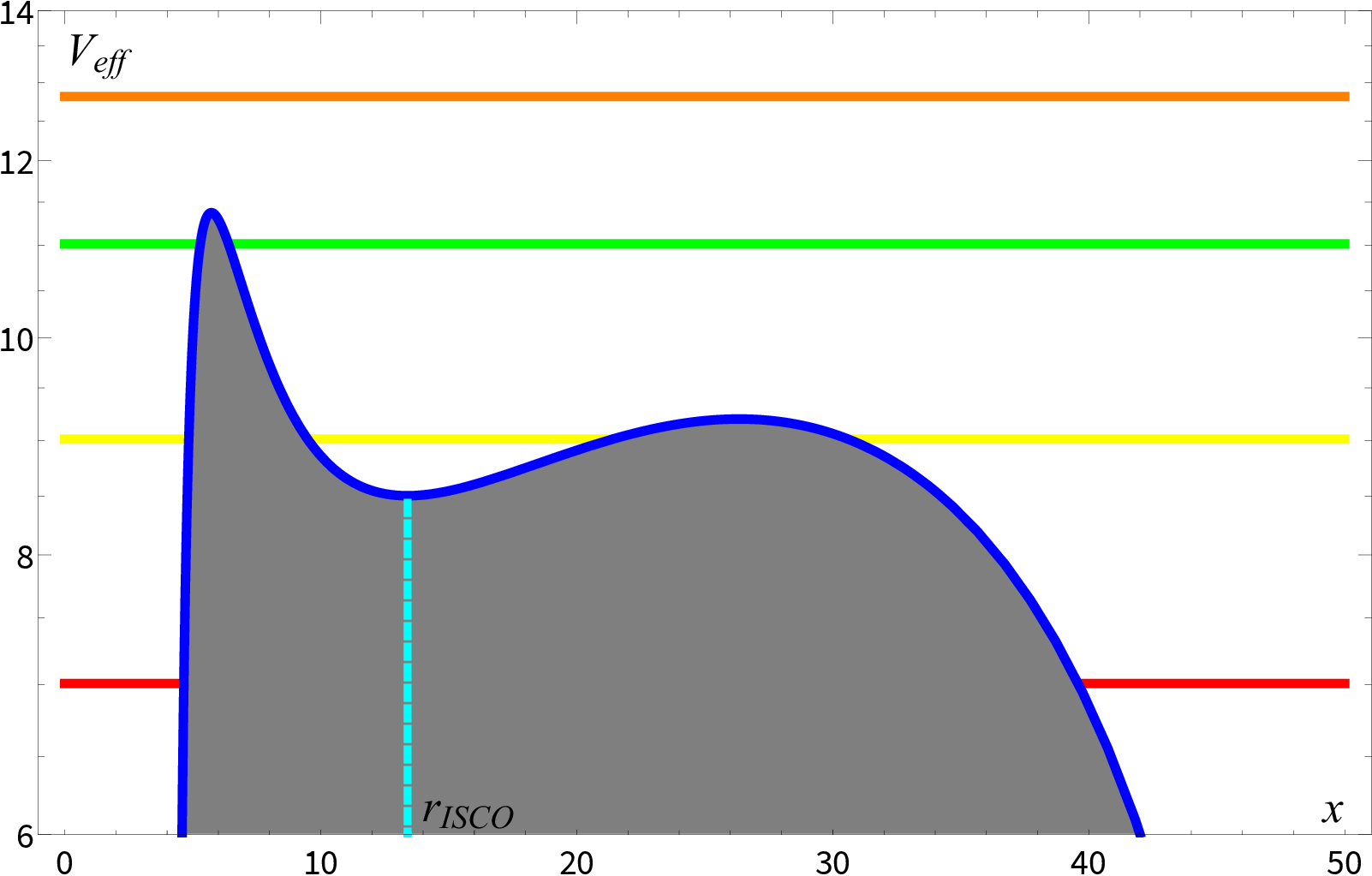}
	\hspace*{0.5cm}
	\includegraphics[height=4 cm,width=5 cm] {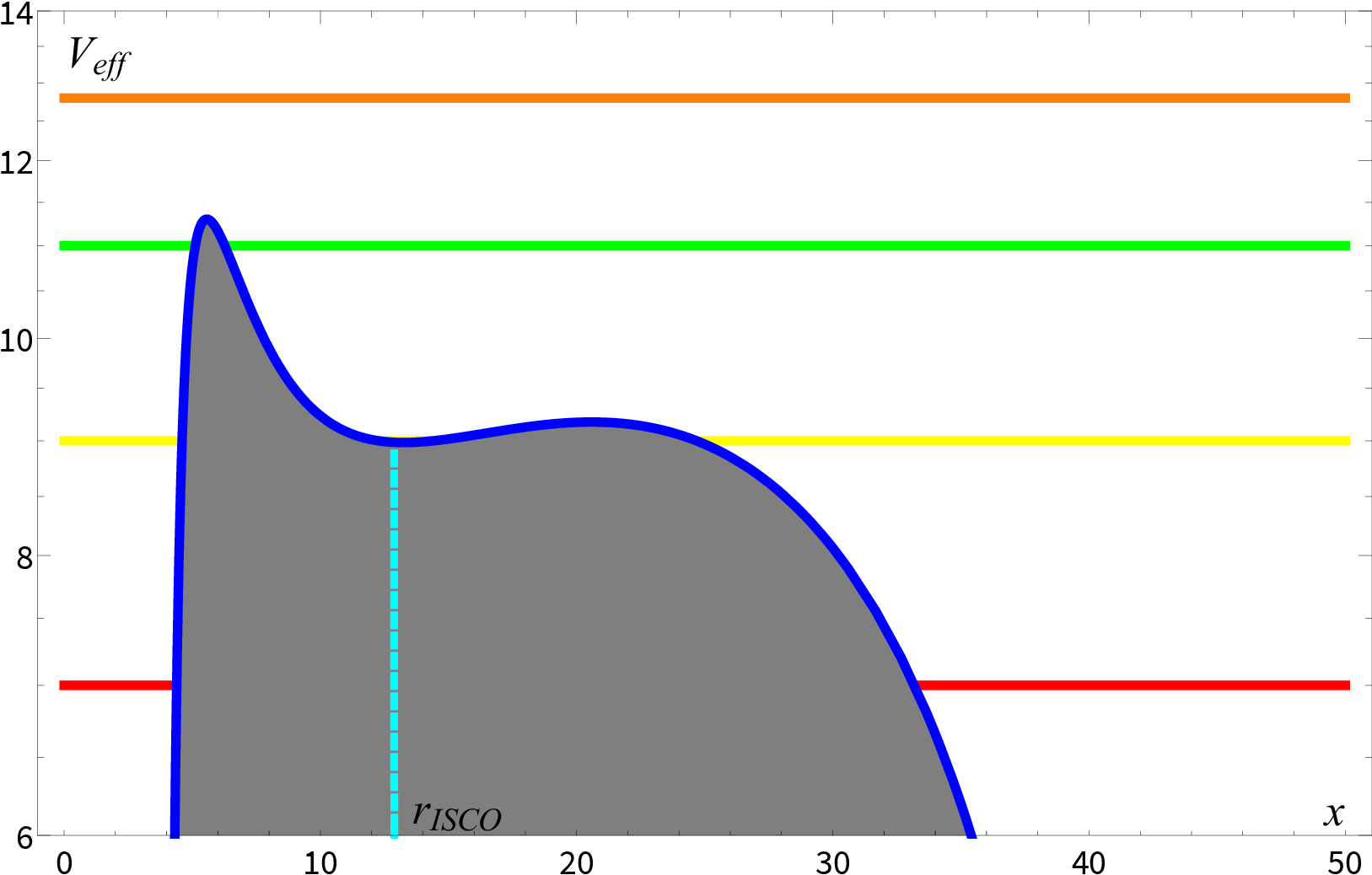}	
	\caption{\scriptsize{The effective potential of the LTB black hole versus distance in three slices of cosmic epochs. Plots have been depicted left to right by $\tau=1.4, 1, 0.6$ from early to late time cosmic epochs. Horizontal lines show possible energy for the particle. Other numerical considerations are the mass of the black hole, $M=1$, the initial angular momentum of the massless particle, $L=10$, $\rho_0=0.0002$, $p_0=0.001$, and $t_{BR}=1$.}}
	\label{figure_2}
\end{figure}

In the following, we analyze the effective potential in some cosmic epochs with numerical methods. So, as a second step, we fix the black hole's mass and the particles' initial angular momentum. About the proper time, we have to notice that we pay attention to some numerical time after forming the black hole horizons while the Big Rip time is equal to 1, $t_{BR}=1$. Therefore, based on our prior research in Ref. \cite{23Es}, these numerical times are considered from $t_1=-0.4$ until $t_2=0.4$. To be precise, two horizons of the LTB black hole have been established before $\tau=1-t_1$, and with passing time these two horizons merge and disappear after $\tau=1-t_2$. Accordingly, we illustrate a general perspective of the effective potential for the cosmological LTB black hole surrounded by a Quintom field in Fig. \ref{figure_1}. On the left-hand plot, we show the behavior of an effective potential versus distance and lifetime of the LTB black hole. Also, on the Right-hand plot of Fig. \ref{figure_1}, we show the behavior of an effective potential versus distance in some time slices. First of all, we recall that the extremum of an effective potential can indicate the possible location of circular orbits so that the minimum corresponds \textbf{stable orbit} while the maximum corresponds an \textbf{unstable orbit} \cite{03Har}. Therefore, based on the illustration of Fig. \ref{figure_1}, we conclude that passing time causes:
\begin{itemize}
	\item{1.} The peak of the potential decreases and shifts to smaller radii. More precisely, in the early universe, after the formation of the black hole, a particle with a given energy is deflected by the black hole's potential and can escape to infinity by following a flyby orbit. In contrast, the same particle with the same energy in late time epochs falls into the black hole.
	\item{2.} The possibility of forming a stable orbit and the Innermost Stable Circular Orbit (ISCO) decreases. We provide further explanation on this issue by presenting several possible particle energies on the potential plot shown in Fig. \ref{figure_2}.
\end{itemize}

\begin{figure}[ht]
	\centering
	
	\includegraphics[height=4 cm,width=4 cm]  {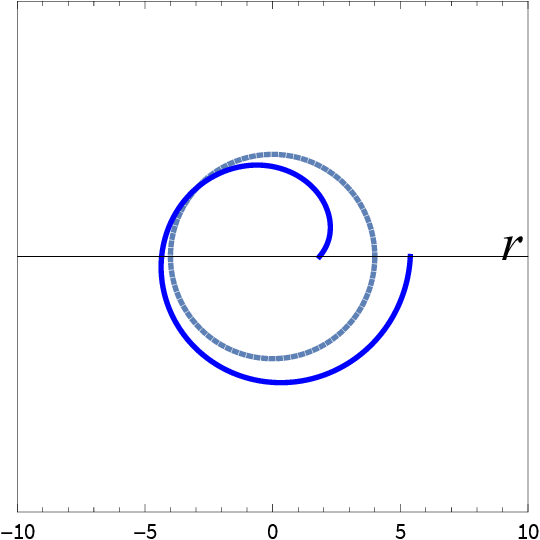}
	\hspace*{1 cm}
	\includegraphics[height=4 cm,width=4 cm] {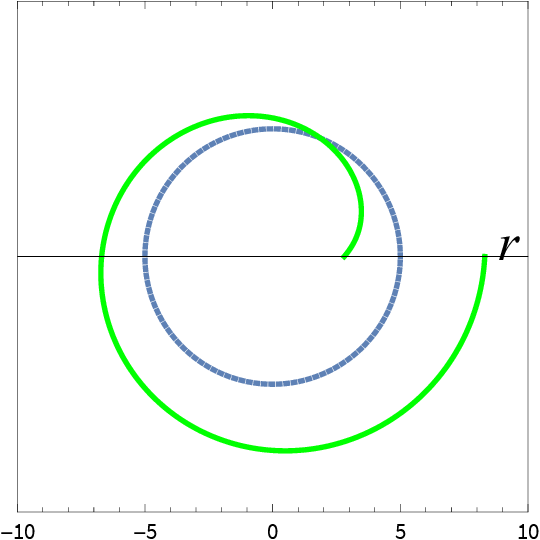}
	\hspace*{1 cm}
	\includegraphics[height=4 cm,width=4 cm]  {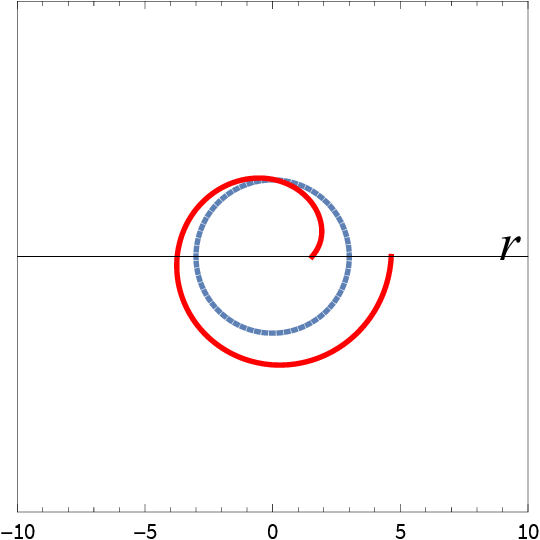}
	\vspace*{1 cm}
	
	\includegraphics[height=4 cm,width=4 cm]  {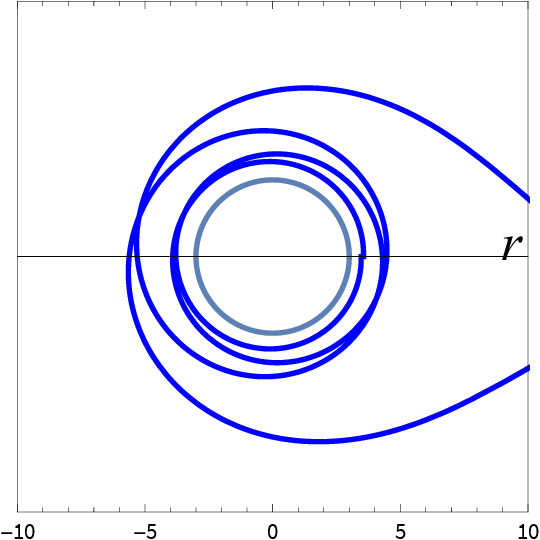}
	\hspace*{1cm}
	\includegraphics[height=4 cm,width=4 cm] {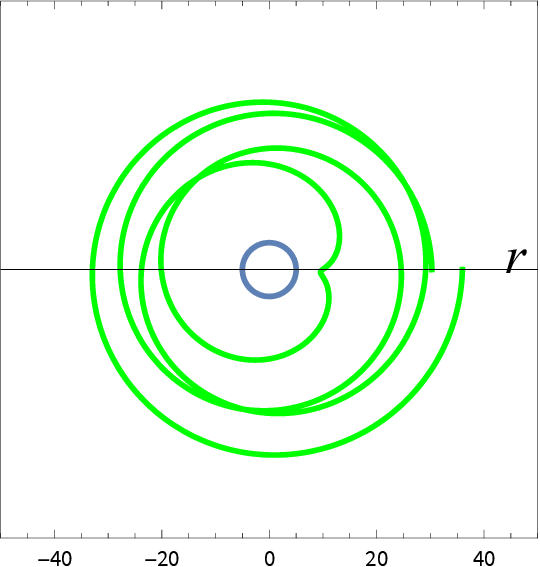}
	\hspace*{1cm}
	\includegraphics[height=4 cm,width=4cm]  {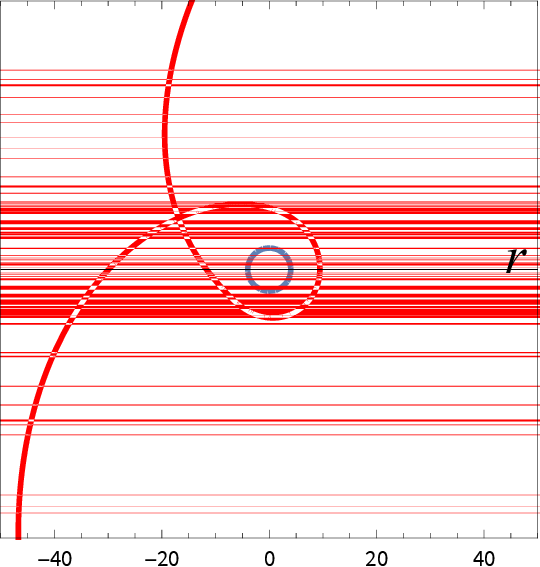}
	\vspace*{1cm}
	
	\includegraphics[height=4 cm,width=4 cm]  {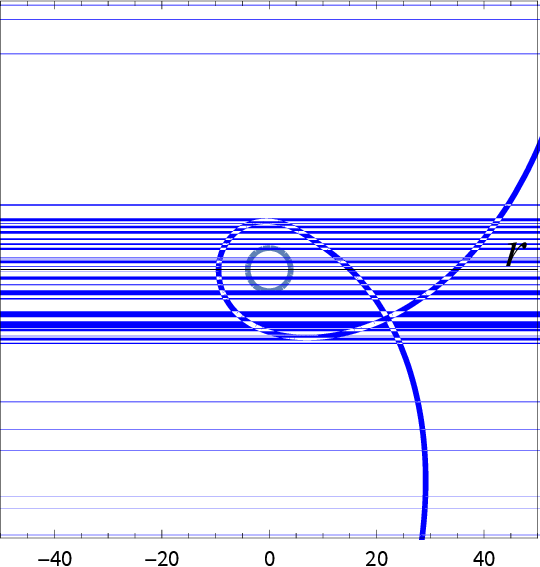}
	\hspace*{1cm}
	\includegraphics[height=4 cm,width=4 cm] {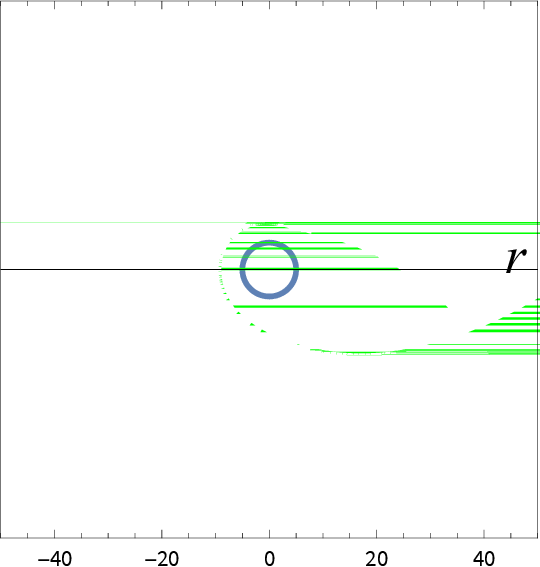}
	\hspace*{1cm}
	\includegraphics[height=4 cm,width=4 cm]  {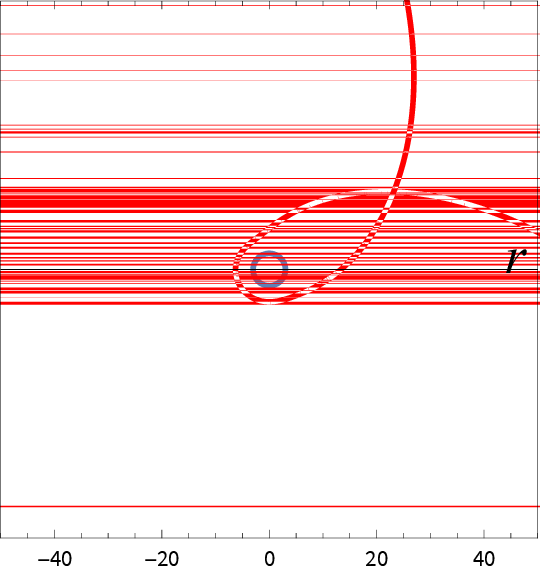}
	
	\vspace*{0.5cm}
	\caption{\scriptsize{Illustration of the possible orbits around the cosmological black hole in some cosmic epochs. From left to right, time evolves from far away towards near the Big Rip singularity. From top to bottom, the energy of the photon increases. The first row illustrates the terminating bound orbit for a low energy photon moves from near the black hole. If the energy of a photon increases, in the second row, the early universe allows for the possibility of forming a stable bound orbit. However, as time progresses toward late time epochs, this possibility diminishes. In this case, a photon can originate from a distant region and orbit the black hole between two turning points. If the energy increases more, a flyby orbit has been expected for all cosmic epochs. Finally, if the photons' energy is more than the peak of the effective potential, there are figures like the first row except this time a particle comes from far away; It falls into the black hole on the terminating escape orbit. Note that dotted blue circle in the center of each plot shows the cosmological horizon of the black hole.}}
	\label{figure_3}
\end{figure}

Plots in Fig. \ref{figure_2}, left to right, demonstrate the passing time from the early universe to late time epochs. Also, the horizontal lines represent the possible energy for the particles based on Eqs. (\ref{E-tot}), (\ref{Mass}) and, (\ref{Sol3}). In this figure, we highlight that, in general, there are two possible initial positions for the particles: they can originate from a distance region—represented by the horizontal lines on the right side of the effective potential—or from near the black hole—represented by the horizontal line on the left side of the effective potential. Based on the particles' energy and the initial positions of the particle, there are four general situations:

\begin{itemize}
\item{1.} 
The red line shows that low-energy particles, originating from near the black hole, are influenced by its potential and eventually fall into it. In this case, the particle follows a \textbf{Terminating Bound Orbit}. On the other hand, if it approaches the black hole from a distant region, it will follow a Flyby Orbit, which will be discussed in item 3.

\item{2.} As the particle's energy increases, it can reach the minimum of the effective potential, allowing it to become trapped in the black hole's gravitational field and orbit in an ISCO. As illustrated in Fig. \ref{figure_2} with the cyan-dashed line, the ISCO radius decreases as cosmic time evolves from early universe to late time epochs. This behavior can be explained by expansion-driven gravitational weakening, where the evolving background of the LTB black hole, influenced by cosmic expansion due to the Quintom field, modifies the effective gravitational potential, causing stable orbits to shift inward over time. Additionally, the time-dependent nature of the black hole itself plays a role, as the LTB black hole evolves within a dynamically changing cosmological background, unlike stationary solutions where the ISCO depends only on mass and spin. The evolution of the apparent horizon further influences the black hole’s gravitational pull, leading to a gradual reduction in the ISCO radius. Furthermore, as the universe approaches the Big Rip singularity, the rapid expansion weakens the gravitational potential, eventually causing stable circular orbits to vanish. At late time epochs, particles near the ISCO either fall into the black hole or escape outward due to the increasing instability of the orbits.

In addition, in the early universe, the formation of stable orbits is more likely; however, as time passes toward late time epochs, this possibility decreases. Therefore, for more energetic photons (represented by the yellow lines) coming from middle regions, two possible scenarios emerge: In the early universe, the particle oscillates between two turning points around the black hole, following a \textbf{Bound Orbit} (see the left and middle second-row plots in Fig. \ref{figure_3}). It is important to note that the turning points are recognized by the intersections of the potential curve and the yellow line in the left and middle plots of Fig. \ref{figure_2}. We emphasize that the likelihood of forming such a Bound Orbit is significantly higher in the early universe. In contrast, in the late time epochs, the particle's trajectory is deflected under the black hole's gravitational attraction, resulting in a Flyby Orbit, a type of scattering orbit (see the right second-row plot in Fig. \ref{figure_3}). Additionally, for a particle originating from a region closer to the black hole, the outcome mirrors the case described in item 1, leading to a Terminating Bound Orbit.

\item{3.} The green line for a particle coming from far away corresponds to a \textbf{Flyby Orbit} (see the third-row plots in Fig. \ref{figure_3}). A flyby orbit occurs when a particle approaches from a distant region, interacts with the black hole’s gravitational field, experiences a change in trajectory, and eventually escapes to infinity. This type of orbit is present in all cosmic epochs; however, its probability increases over time as the likelihood of forming bound orbits decreases. This type of orbit is present for all particles with energy less than the potential peak approaching the black hole from a distant region in all cosmic epochs; however, its probability increases over time as the likelihood of forming bound orbits decreases. This issue can be understood by considering an evolution of the effective potential. As time passes and the universe expands due to the Quintom field, the gravitational potential weakens, and the maximum of the effective potential shifts inward. As a result, more energetic particles that might have been captured in stable orbits in early universe are more likely to follow flyby orbit instead. Simultaneously, the shrinking of ISCO region reduces the stability of bound orbits, reinforcing the rise in flyby orbits. On the other hand, as the universe nears the Big Rip singularity, rapid expansion eliminates stable circular orbits, forcing particles near the ISCO to either fall into the black hole or escape to infinity.\\
Additionally, the deflection angle of flyby orbits is an important factor for observational studies, as it could provide insights into gravitational lensing effects and black hole interactions with surrounding particles. Similar to the previous cases, a Terminating Bound Orbit is expected for a particle originating from a region close to the black hole.

\item{4.}
The energy of a particle can increase again until it exactly reaches the peak of an effective potential so that it goes around the central mass at the unstable orbit. Suppose the particles' energy is more than the maximum of an effective potential. In the case, an orange line, the particle comes from a distance region, follows a \textbf{Terminating Escape Orbit} and falls into the black hole.
\end{itemize}

To visualize the shape of the mentioned orbits, we proceed with numerical methods. First, we employ the metric (\ref{LElement}) for a massless particle ($ds^2=0$) moving in the equatorial plane ($\theta=\pi/2$). Consequently, we obtain

\begin{equation}\label{chi}
	\Big[\frac{dr}{d\varphi}\Big]^2=\frac{e^{2\phi}}{L^2}\Bigg[\frac{1}{e^{\bar{\phi}}} E^2- \frac{e^{\phi}}{e^{\bar{\phi}}}\Big(\frac{L}{e^{\phi}}\Big)^2 \Bigg]\equiv\chi,
\end{equation}
where $E=[\frac{dt}{ds}]$, and $L=e^\phi [\frac{d\varphi}{ds}]$. We substitute Eqs. (\ref{Sol3}) and (\ref{Sol2}) in Eq. (\ref{chi}), and we obtain a differential equation in terms of $r,\varphi, L,$ and $E$ as the following form

\begin{equation}\label{chi2}
	d\varphi=\frac{dr}{\sqrt{\chi}},
\end{equation}

Next, we put $\tau=1.4, 1, 0.6$ separately in Eq. (\ref{chi2}) and gain three equations. Afterward, substituting $M=1$, $L=10$, $\rho_0=0.0002$, $p_0=0.001$ we have three numerical differential equations. Finally, Based on Fig. \ref{figure_2}, we consider different energy for photons and illustrate the photon's trajectory in Fig. \ref{figure_3}.

If a low-energy particle is located near the black hole (showed by the red line in Fig. \ref{figure_2}), in all cosmic epochs, it falls into the black hole through the terminating bound orbit as we show them in the first row in Fig. \ref{figure_3}. Also, this fate is expected for the more energetic particles that we can see at the left side of the effective potential plots as the yellow and green lines. As the energy increases from the red to yellow lines in Fig. \ref{figure_2}, there is a possibility of forming a stable orbit around the black hole in the early cosmic epochs. Accordingly, we illustrate a stable bound orbit in the left second-row plot of Fig. \ref{figure_3}. However, as time progresses toward the Big Rip singularity, the likelihood of stable orbit formation decreases. Consequently, we depict a flyby orbit for a particle originating from a distant region in the right second-row plot of Fig. \ref{figure_3}. As the energy increases, the black hole can only alter the particle's trajectory, resulting in a flyby orbit, as illustrated in the third row of Fig. \ref{figure_3}. The final possibility is the terminating escape orbit, which occurs for particles with energy exceeding the peak of the effective potential at all cosmic epochs. In this case, the trajectory resembles those in the first row of Fig. \ref{figure_3}, except that the more energetic particle originates from a distant region and is ultimately absorbed by the black hole. In all plots of Fig. \ref{figure_3}, the dotted blue circle represents the largest horizon of the black hole, known as the cosmological horizon. As discussed in Section II, we observe that its size evolves due to the presence of the Quintom field as part of the universe's content.

\subsection{Stability of the Orbits}
Based on our understanding of black hole potential plots, the minimum point of the effective potential curve indicates the possible location of a stable circular orbit at a specific radius around the black hole. This minimum corresponds to a stable balance where a particle can maintain a circular trajectory under the influence of the black hole’s gravitational field. In contrast, the maximum of the potential represents an unstable equilibrium, where small perturbations will cause the particle to either fall into the black hole or escape outward. We concluded in the previous subsection that, it can be inferred that if time is the only variable, once a circular orbit is formed in the early universe, its minimum radius gradually decreases as the black hole evolves in an accelerating universe toward the Big Rip singularity.\\ 
To further investigate the nature of stable orbits at a specific cosmic epoch near the early universe, we fix time to $\tau=1.4$ and analyze the effective potential numerically using Eq. (\ref{veff}). Our results show that within a specified range, $8<L<12$, the general behavior of the potential remains preserved.A key finding is that as the angular momentum increases, the radius of ISCO decreases, while the radius corresponding to the maximum of the potential increases. This means that for a particle with higher angular momentum, the distance between the two turning points—which define the region in which the particle follows a Bound Orbit around the black hole—becomes larger. Consequently, larger angular momentum allows for a wider range of bound motion, influencing the possible paths of particles in the black hole’s gravitational field.\\
Additionally, since the potential values increase with increasing $L$, more energetic photons can be captured by the black hole, forming circular or general Bound Orbits. On the other hand, as expected, with increasing angular momentum, the two peaks of the potential move closer to each other. This behavior reduces the probability of Flyby Orbit formation and increases the likelihood of direct particle fall into the black hole.\\
\begin{figure}[ht]
	\centering
	\includegraphics [height=7 cm,width=9 cm] {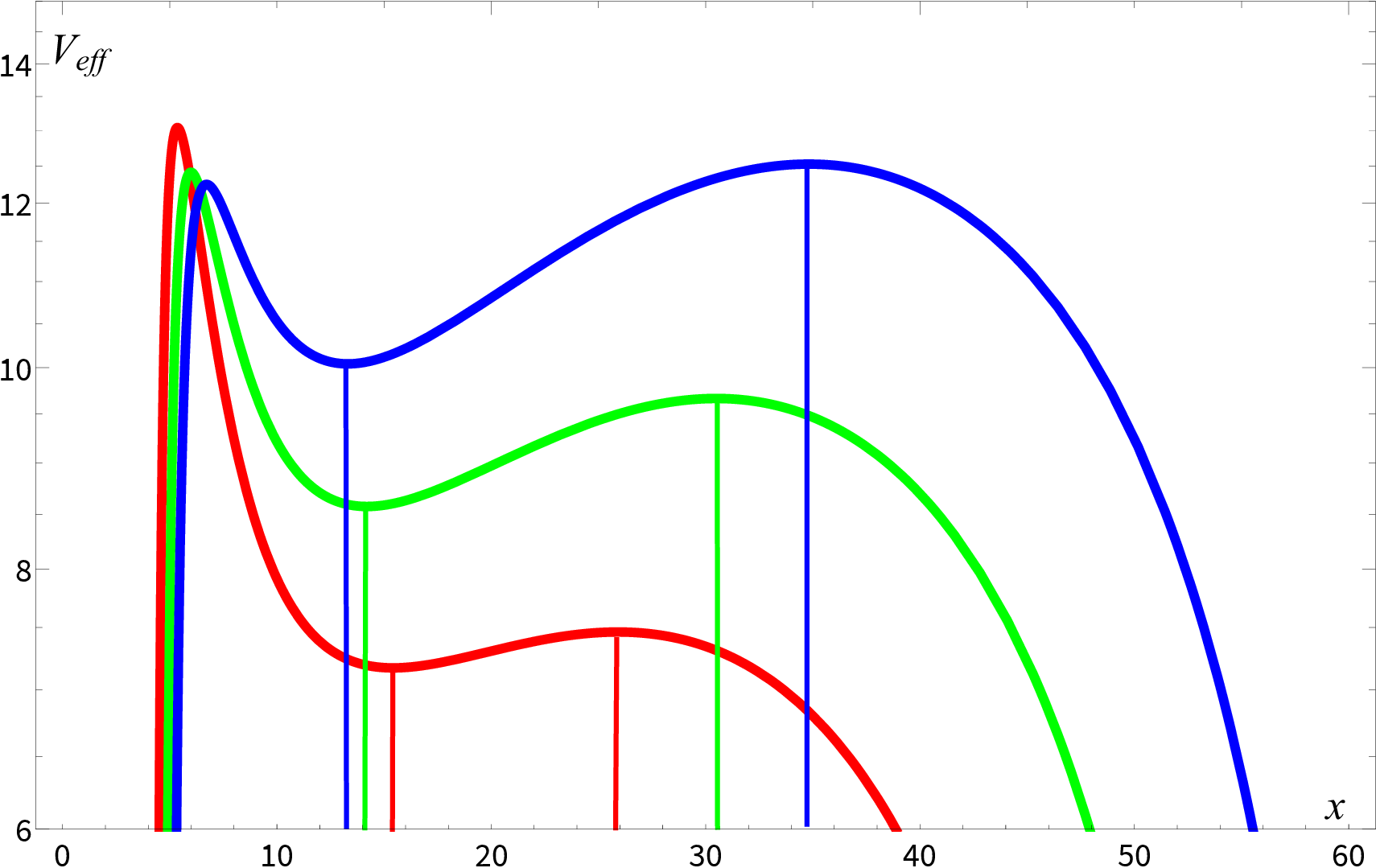}
	\caption{\scriptsize{The behavior of the Effective Potential versus distance. Plots have been depicted with $L=9,10,11$ red to blue and, fixed mass $M=1$, while $\rho_0=0.0002$, $p_0=0.001$, and $\tau=1.4$.}}
	\label{figure_4}
\end{figure}

\subsection{Outlook}
In this study, we have demonstrated the evolution of particle orbits around an LTB black hole in an accelerating universe, from early to late-time epochs. Previous investigations of geodesics in McVittie spacetimes have explored the impact of cosmic expansion on orbital stability. Antoniou and Perivolaropoulos in Ref. \cite{16Ant} examined the dissolution of bound orbits in a phantom energy driven universe, calculating the dissociation times. Nolan \cite{14Nol} classified stable orbits in McVittie backgrounds, demonstrating that bound motion can persist under certain conditions, with the ISCO evolving asymptotically like Schwarzschild and Schwarzschild-de Sitter spacetimes. Baker \cite{02Bak} introduced the concept of attraction domains, showing that cosmic expansion defines gravitational bound regions, which eventually disband as acceleration increases. Our study expands upon these findings by analyzing an LTB black hole with a Quintom field, allowing us to track the time-dependent evolution of orbital structures. Unlike McVittie spacetimes, where the black hole’s mass remains constant, our model features an inhomogeneous expanding background, leading to a shrinking ISCO and an increasing dominance of flyby orbits. While previous studies have primarily focused on static classifications, dissociation times, or large-scale gravitational structures, we investigate how the effective potential evolves over time, influencing the transition between different orbital types. Both our study and previous work on McVittie geodesics \cite{16Ant} confirm that a phantom energy driven Big Rip leads to the breakdown of stable orbits. However, we explicitly demonstrate that as the Big Rip approaches, stable orbits vanish, and flyby orbits become dominant, marking a transition from bound to unbound motion due to the expansion driven weakening of gravitational forces. This result offers a deeper understanding of how cosmic evolution affects orbital dynamics in black hole spacetimes.\\
A natural extension of this work is to investigate the evolution of orbits in a Kerr-Newman black hole in an accelerated universe, as outlined in our previous studies \cite{16Sor}. In particular, understanding orbit stability and deflection angles in such spacetimes could have significant implications for the observation of early galaxies and X-ray sources. The stability of orbits and the deflection of light around black holes are directly related to key observational signatures, such as gravitational lensing and the behavior of light in the vicinity of supermassive black holes. If the evolution of the effective potential in a Kerr black hole can be analyzed across different cosmic epochs, we can better estimate particle trajectories and investigate the transition of orbital properties over time.\\
Furthermore, the shrinking ISCO in an accelerating universe may lead to observable effects that could be tested with future astrophysical surveys. A key consequence is the potential modification of gravitational lensing and black hole shadow properties, as the ISCO directly influences the photon sphere and light deflection around the black hole. Detecting subtle shifts in lensing effects or deviations in black hole shadows over cosmic time could provide empirical evidence of evolving spacetime backgrounds. Additionally, since the ISCO defines an inner edge of accretion disks, its contraction could lead to hotter, more compact accretion structures, altering the X-ray spectra emitted by black hole systems. High precision X-ray observations, particularly with next generation instruments designed for black hole accretion studies in cosmological contexts, could detect such changes.\\
Moreover, quasi-periodic oscillations (QPOs) in X-ray binaries, which are closely linked to ISCO properties, may offer a direct observational probe of geodesic shifts in time-dependent black hole spacetimes. The time evolution of QPO frequencies over cosmic history could provide valuable insights into the dynamical effects of cosmic expansion on black hole orbits. Future observations using cutting-edge instruments such as the Event Horizon Telescope (EHT), the Athena X-ray Observatory, and gravitational wave detectors could offer crucial insights into how cosmic expansion influences the structure and dynamics of black hole environments.\\
By further exploring these theoretical and observational implications, we can deepen our understanding of how accelerating expansion influences geodesic motion, gravitational lensing, and black hole evolution, bridging the gap between general relativity, astrophysics, and cosmology.
\section{Discussion and Conclusion}
In this study, we analyzed the orbital evolution of massless particles around an LTB black hole in an accelerating universe driven by the Quintom field. Indeed, the Quintom field also provided a fine description for crossing $-1$ of the equation of state, and the LTB black hole provided a nice description of the dynamical black hole in the accelerated universe. Unlike stationary black holes, the time-dependent nature of the metric significantly influences the stability and classification of geodesic motion. By deriving the effective potential for such a black hole in Eq. (\ref{veff}) and analyzing, we demonstrated that both the peak and minimum of the potential shift over time from the early universe until the late time cosmic epochs, altering the landscape of possible orbits.

Moreover, the behavior of an ISCO as a crucial characteristic of black hole spacetimes, significantly impacting both the geodesic structure and observational features of the black hole have been investigated. A key result of our finding is that as the universe expands, stable circular orbits become less probable. Our numerical analysis confirmed that the ISCO radius decreases over cosmic time, leading to a higher frequency of flyby orbits and a lower probability of stable bound orbits. This transition is a direct consequence of expansion-driven gravitational weakening, causing orbits that were once stable in the early universe to become unstable or scattered by the black hole's gravitational field in later epochs.

Additionally, we investigated the role of angular momentum in shaping the orbital structures. We found that as angular momentum increases, the ISCO contracts while the peak of the effective potential shifts outward. This indicates that higher angular momentum allows for extended bound motion, increasing the likelihood of particle capture by the black hole. However, at very high angular momentum, flyby orbits become less probable, and direct fall into the black hole becomes more dominant.

The implications of our findings extend beyond theoretical modeling. The shrinking ISCO may lead to observable effects, including modifications in gravitational lensing, changes in accretion disk structures, and shifts in QPO frequencies over cosmic time. Future high-precision X-ray telescopes and gravitational wave detectors could provide observational evidence for the time-dependent evolution of ISCO and black hole geodesics in an accelerating universe.

Future research could explore more complex black hole spacetimes, such as Kerr-Newman black holes in an expanding background, where rotation and charge introduce additional degrees of freedom affecting orbital stability. Investigating how these effects modify deflection angles, light propagation, and accretion processes could provide deeper insights into the connection between cosmic expansion and black hole environments. By extending our analysis to observational tests, we may further bridge the gap between general relativity, astrophysics, and cosmology.\\

\textbf {Acknowledgment}\\
We would like to appreciate Kourosh Nozari for helpful discussion and insightful comments on the original draft of this manuscript.

\end{document}